\documentclass[runningheads]{llncs}
\usepackage{graphicx}
\usepackage{booktabs}
\usepackage{graphicx}
\usepackage{todonotes}
\usepackage{balance}       
\usepackage{graphics}      
\usepackage{color}
\usepackage{todonotes}
\usepackage{hyperref}
\usepackage{csquotes}
\usepackage{changes}
\usepackage{tabularx}
\usepackage{booktabs}
\usepackage{textcomp}
\usepackage[skip=3pt,font=small,labelfont= bf]{caption}
\usepackage{subcaption}
\usepackage{makecell}
\usepackage{colortbl}
\usepackage{xcolor}
\usepackage{graphicx}
\usepackage{wrapfig}
\usepackage{booktabs}
\usepackage{sansmathfonts}
\newcommand{\RNum}[1]{\lowercase\expandafter{\romannumeral #1\relax}}

\usepackage [english]{babel}
\MakeOuterQuote{"}
\begin{document}
\title{Human-centered Explainable AI: Towards a Reflective Sociotechnical Approach}
\titlerunning{Human-centered Explainable AI}

%
%
%
\author{Upol Ehsan \and
Mark O. Riedl}
%
%
\authorrunning{Ehsan \& Riedl}
%
\institute{Georgia Institute of Technology\\
Atlanta, GA 30308, USA\\
\email{ehsanu@gatech.edu, riedl@cc.gatech.edu}}
%
%
%
%
%
\maketitle              
\begin{abstract}

Explanations---a form of post-hoc interpretability---play an instrumental role in making systems accessible as AI continues to proliferate complex and sensitive sociotechnical systems. 
In this paper, we introduce Human-centered Explainable AI (HCXAI) as an approach that puts the human at the center of technology design.
It develops a holistic understanding of \textit{``who'' the human is} by considering the interplay of values, interpersonal dynamics, and the socially situated nature of AI systems. 
In particular, we advocate for a \textit{reflective sociotechnical} approach.
We illustrate HCXAI through a case study of an explanation system for non-technical end-users that shows how technical advancements and the understanding of human factors co-evolve.
Building on the case study, we lay out open research questions pertaining to further refining our understanding of ``who'' the human is and extending beyond 1-to-1 human-computer interactions.
Finally, we propose that a \textit{reflective} HCXAI paradigm---mediated through the perspective of Critical Technical Practice and supplemented with strategies from HCI, such as value-sensitive design and participatory design---not only helps us understand our intellectual blind spots, but it can also open up new design and research spaces.
%
%
%
\end{abstract}
\section{Introduction}
From healthcare to finances, human resources to immigration services, many powerful yet ``black-boxed'' Artificial Intelligence (AI) systems have been deployed in consequential settings.
This ubiquitous deployment creates an acute need to make AI systems understandable and explainable \cite{barocas2016big,berk2012criminal,bermingham2011using,chen2012business,galindo2000credit}.
{\em Explainable} AI (XAI) refers to artificial intelligence and machine learning techniques that can provide human-understandable justification for their output behavior. 
Much of the previous and current work on explainable AI has focused on {\em interpretability}, which we view as a property of machine-learned models that dictates the degree to which a human user---AI expert or non-expert user---can come to conclusions about the performance of the model given specific inputs. 
{\em Explanation generation}, on the other hand, can be described as a form of post-hoc interpretability \cite{2016arXiv160603490L,miller2017explanation,ribeiro2016should,yosinski2015understanding}. 
An important distinction between interpretability and explanation generation is that explanation does not necessarily elucidate precisely how a model works but aims to provide useful information for practitioners and  users in an accessible manner.

While the letters ``HCI'' might not appear in ``XAI'', explainability in AI is as much of a Human-Computer Interaction (HCI) problem as it is an AI problem, if not more. 
Yet, the human side of the equation is often lost in the technical discourse of XAI. 
Implicit in Explainable AI is the question: ``explainable to whom?'' 
In fact, the challenges of designing and evaluating ``black-boxed'' AI systems depends crucially on \textit{``who''} the human in the loop is. 
Understanding the \textit{``who''} is crucial because it governs what the explanation requirements for a given problem. 
It also scopes \textit{how} the data is collected, \textit{what} data can be collected, and the most effective way of describing the \textit{why} behind an action.
For instance: with self-driving cars, the engineer may have different requirements of explainability than the rider in that car.
As we move from AI to XAI and recenter our focus on the human---through Human-centered XAI (HCXAI)---the need to refine our understanding of the \textit{``who''} increases. 
As the domain of HCXAI evolves, so must our epistemological stances and methodological approaches. 
Consequential technological systems, from law enforcement to healthcare, are almost always embedded in a rich tapestry of social relationships. 
If we ignore the socially situated nature of our technical systems, we will only get a partial and unsatisfying picture of the \textit{``who''}.  

In this paper, we focus on unpacking \textit{"who" the human is} in Human-centered Explainable AI and advocate for a
sociotechnical approach. 
We argue that, in order to holistically understand the socially situated nature of XAI systems, we need to incorporate both social and technical elements.
This sociotechnical approach can help us critically reflect or contemplate on implicit or unconscious values embedded in computing practices so that we can understand our epistemological blind spots. 
Such contemplation---or reflection---can bring unconscious or implicit values and practices to conscious awareness, making them actionable. 
As a result, we can design and evaluate technology in a way that is sensitive to the values of both designers and stakeholders. 

We begin by using a case study in Section 2, to delineate how the two strands of HCXAI---technological development and the understanding of human factors---evolve together.
The case study focuses on both the technological development and the human factors of how non-expert users perceive different styles of automatically-generated rationales from an AI agent~\cite{ehsan2017rationalization,ehsan2019automated}. 
In Section 3, using the insights from the study, we share future research directions that demand a sociotechnical lens of study. 
Finally, in Section 4, we introduce the notion of a \textit{Reflective} HCXAI paradigm and outline how it facilitates the sociotechnical stance. 
We overview related concepts, share strategies, and contextualize them by using scenarios. 
We conclude by delineating the challenges of a reflective approach and presenting a call-to-action to the research community.

\section{Case Study: Rationale Generation}

The case study is based on our approach of post-hoc explanation generation called \textit{rationale generation}, a process of producing a natural language rationale for agent behavior as if a human had performed the behavior and verbalized their inner monologue (for details, please refer to our papers \cite{ehsan2017rationalization,ehsan2019automated}). 
The main goal for this section is to highlight the meta-narrative of our HCXAI journey; in particular, how the two processes---technological development in XAI and understanding of human-factors---co-evolve. 
Specifically, we will see how our understanding of human factors improve over time. 

As an analogy while we go through the two phases of the case study, consider a low-resolution picture, say $16\times16$ pixels, that gets updated to a higher resolution photo, say $256\times256$ pixels, of the same subject matter. 
Not only does better technology (in our analogy, a better camera) afford a higher resolution image, but the high-resolution image also captures details previously undetectable, which, when detected broadens our perspective and facilitates new areas of interest. 
For instance, we might want to zoom in on a particular part of the picture that requires a different sensor. 
Had we not been able to broaden our perspective and incorporate things previously undetectable, we would not have realized the technical needs for a future sensor. 
As we can see, the two things---the camera technology and our perspective of the subject matter--- build on each other and co-evolve. 
For the rest of the section, we will provide a brief overview of \textit{rationale generation}, especially its technical and philosophical underpinnings. Finally, we will share key takeaways from the two phases of the case study. For fine-grained empirical details, please refer to \cite{ehsan2017rationalization,ehsan2019automated}.

With this narrative of co-evolution in mind, let us look at the philosophical and technical intuitions behind rationale generation.
The philosophical intuition behind rationale generation is that humans can engage in effective communication by verbalizing plausible motivations for their actions, even when the verbalized reasoning does not have a consciously-accessible neural correlate of the decision-making process \cite{fodor1994elm,block2007consciousness,block2005two}. 
Whereas an explanation can be in any communication modality,  we view {\em rationales} as natural language explanations. Natural language is arguably the most accessible modality of explanation. 
However, since rationales are natural language explanations, there is a level of abstraction between the words that are generated and the inner workings of an intelligent system. 
This motivates a range of research questions pertaining to how the choice of words for the generated rationale affect human factors such as \textit{confidence} in the agent’s decision, \textit{understandability}, \textit{human-likeness}, \textit{explanatory power}, \textit{tolerance to failure}, and perceived \textit{intelligence}. 

\begin{figure}[t]
    \vspace{-5mm}
    \begin{center}
        \includegraphics[width=0.5\linewidth]{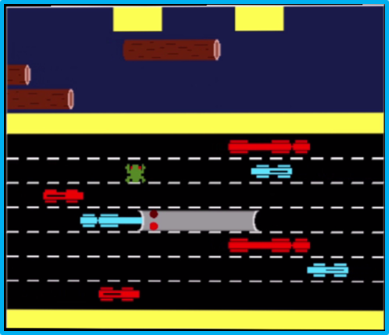} 
    \end{center}
    \captionsetup{width=0.95\linewidth, font= small}
    \caption{A screenshot of the game {\em Frogger}. The green frog \textit{Frogger}, seen in the middle of the image, wins if it can successfully reach the goal (yellow landing spots) at the top of the screen.}
    \label{fig:Game_screenshot}
    \vspace{-5mm}
\end{figure}

From a technical perspective, \textit{rationale generation} is treated as the problem of translating the internal state and action representations into natural language using computational methods. 
It is fast, sacrificing an accurate view of the agent's decision-making process for a real-time response, making it appropriate for real-time human-agent collaboration \cite{ehsan2019automated}. 
In our case study, we use a deep neural network trained on human explanations---specifically a neural machine translation approach~\cite{luong2015effective}---to explain the decisions of an AI agent that plays the game of {\em Frogger}.
In the game, {\em Frogger} (the frog, controlled by the player) has to avoid traffic and hop on logs to cross the river in order to reach its goal at the top of the screen, shown in Figure~\ref{fig:Game_screenshot}.
{\em Frogger} can be thought of as a gamified abstraction of a sequential decision-making task, requiring the player to think ahead in order to choose a good action.
Furthermore, sequential tasks are typically overlooked in explainable AI research.
We trained a reinforcement learning algorithm to play the game, not because it was difficult for the AI to play but because reinforcement learning algorithms are non-intuitive to non-experts, even though the game is simple enough for people to learn and apply their own intuitions. 

Having contextualized the approach, we will break the case study into two main phases. For ease of comparison in the co-evolution, we will cover the same topics for both phases---namely, data collection and corpus creation, neural translation model configuration, and evaluation. Table \ref{tab:two-phase-comparison}, at the end of this section, summarizes each aspect and provides a side-by-side comparison of the two phases. 
%

\subsection{Phase 1: Technological Feasibility \& Baseline Plausibility}
In the first stage of the project \cite{ehsan2017rationalization}, our goal was an existence proof---to show that we could generate satisfactory rationales, treating the problem of explanation generation as a translation problem. At this stage, the picture of the human or end-user was not well-defined by construction because we did not even have the technology to probe and understand them. 
\subsubsection{Data Collection and Corpus Creation}
There is no readily-available dataset for the task of learning to generate explanations. 
Thus, we had to create one.
We developed a methodology to remotely collect live ``think-aloud'' data from players as they played through a game of {\em Frogger} (our sequential environment). 
To get a corpus of coordinated game states, actions, and explanations, we built a modified version of {\em Frogger} in which players simultaneously play the game and explain each of their actions. 

In the first phase, 12 participants provided a total of 225 action-rationale pairs of gameplay. 
To create a training corpus appropriate for the neural network, we used these action-rationale annotations to construct a grammar for procedurally-generating synthetic sentences, grounded in natural language. 
This grammar used a set of rules based on in-game behavior of the {\em Frogger} agent to generate rationales that resemble the crowd-sourced data previously gathered. 
This entails that our corpus for Phase 1 was semi-synthetic in that it contained both natural and synthetic action-rationale pairs. 
\subsubsection{Neural Model Configuration}

We use a 2-layered encoder-decoder recurrent neural network (RNN) \cite{bahdanau2014neural,luong2015effective} with attention to teach our network to generate relevant natural language explanations for any given action (for details, see \cite{ehsan2017rationalization}). 
These kinds of networks are commonly used for machine translation tasks (translating from one natural language to another), but their ability to understand sequential dependencies between the input and the output make them suitable for explanation generation in sequential domains as well. 

Empirically, we found that a limited, $7\times7$, window for observation around a reinforcement learning agent using tabular Q-learning \cite{watkins92} leads to effective gameplay. 
We gave the rationale generator the same $7\times7$ observation window that the agent needs to learn to play. We refer to this configuration of the rationale generator as the \textit{focused-view} generator.
\subsubsection{Evaluation}
For this phase, the evaluation was part procedural- and part human-based. 
For the procedural evaluation, we used BLEU~\cite{papineni2002bleu} scores--a metric often used in machine translation tasks--with a 0.7 accuracy cutoff. 
Since the grammar contained rules that govern when certain rationales are generated, it allowed us to compare automatically-generated rationales against a ground truth. 
We found that our approach significantly outperformed both rationales generated by a random model and a majority classifier for environments with different obstacle densities~\cite{ehsan2017rationalization}.

With the accuracy of the rationales established via procedural evaluation, we needed to see if these rationales were satisfactory from a human-centered perspective. 
On the human evaluation side, we used a mixed-methods approach where 53 participants watched videos of 3 AI agents explaining their actions in different styles. 
After watching the videos, participants ranked their satisfaction with the rationales given by each of the three agents and justified their choice in their own words. 
We found that our system produced rationales with the highest level of user-satisfaction. 
Qualitative analysis also revealed important components of “satisfaction” such as {\em explanatory power} that were important for participants' confidence in the agent, the rationale's perceived relatability (or humanlike-ness), and understandability. 

To summarize, the goal of this first phase was an existence proof of the technical feasibility of generating rationales.
We learned that our neural machine translation approach produced accurate rationales and that humans found them satisfactory.
Not only did this phase inspire us to build on the technical side, but the understanding of the human factors also helped us design better human-based evaluations for the next phase. 
%
%
%
%
%
%
\subsection{Phase 2: Technological Evolution \& Human-centered Plausibility}
Phase 2~\cite{ehsan2019automated} is about taking the training wheels off and making the XAI system more human-centered. Everything builds on our learnings from Phase 1. 
Here, you will see how the data collection and corpus is human-centered and non-synthetic, how our network produces two styles of rationales, and how our evaluation was entirely human-based. 
\begin{figure}[t]
    \centering
    \begin{minipage}[t]{0.35\textwidth}
        \centering
        \includegraphics[width=\textwidth]{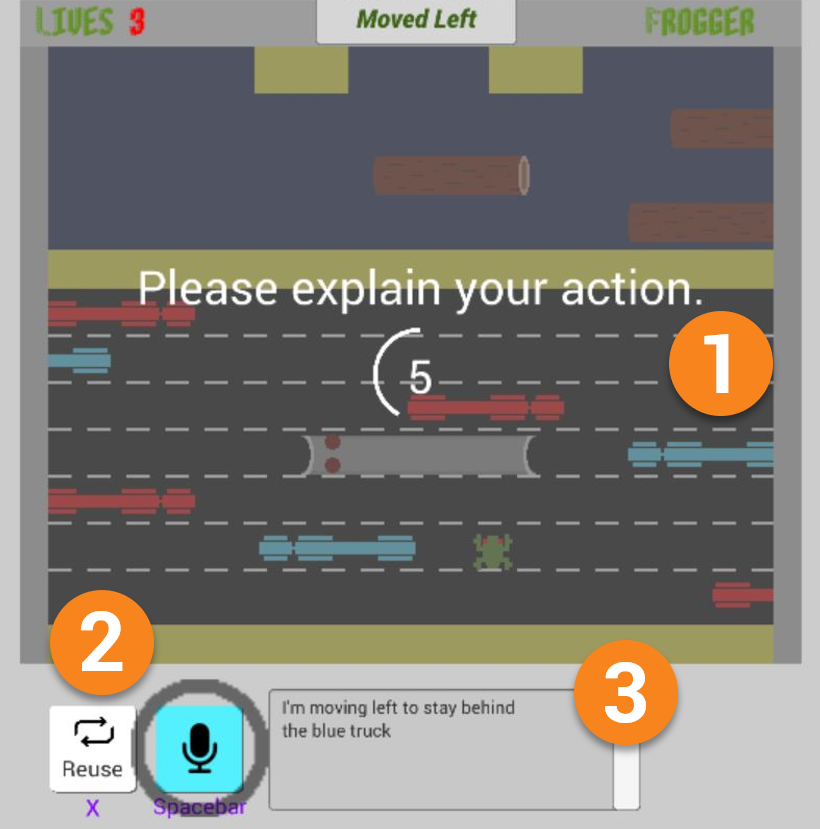} 
        \captionsetup{width=0.95\linewidth, font= small}
        \caption{The rationale collection process. (1)~Game pauses after each action. (2)~Automated speech recognition transcribes the rationale. (3)~Participants can view and edit the transcribed rationales.}
        \label{fig:Play_game}
    \end{minipage}\hfill
    \begin{minipage}[t]{0.6\textwidth}
        \centering
        \includegraphics[width=\textwidth]{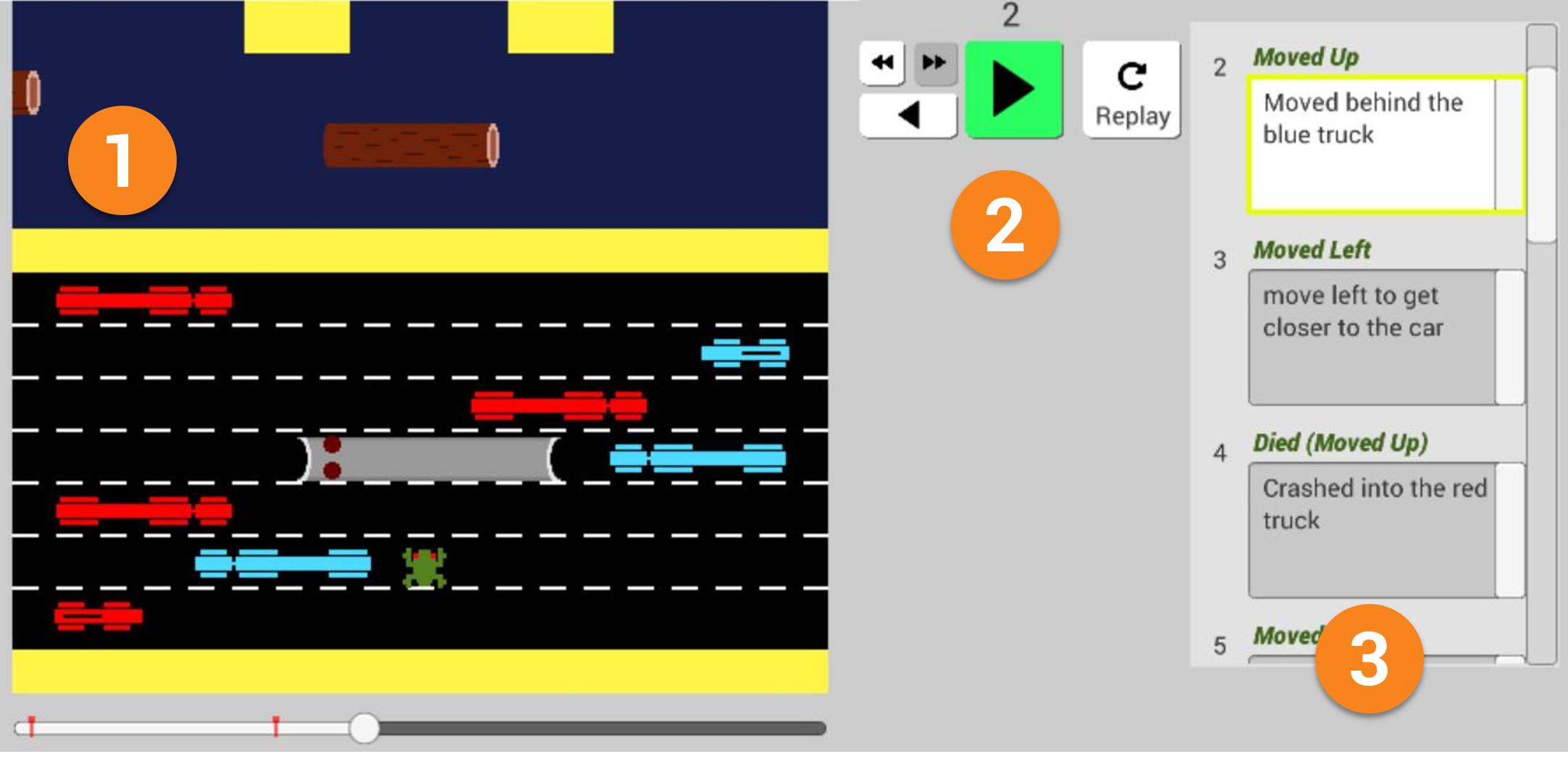} 
        \captionsetup{width=0.95\linewidth, font = small}
        \caption{The rationale review process where players can step-through each of their action-rationale pairs and edit if necessary. (1)~Players can watch an action-replay while editing rationales. (2)~Buttons control the flow of the step-through process. (3)~Rationale for the current action gets highlighted for review.}
        \label{fig:Review_rationale}
    \end{minipage}
\end{figure}
\subsubsection{Data Collection and Corpus Creation}

We expanded the data collection paradigm introduced in phase 1. 
For phase 2, we built another modified version of {\em Frogger} that facilitates a human-centered approach and generates a corpus that is entirely natural-language--based (no synthetic-grammar--generated sentences). 
We split the data collection into three phases: 
$(1)$~a guided tutorial, 
$(2)$~rationale collection, and 
$(3)$~transcribed explanation review. 
The guided tutorial ensured that users are familiar with the interface and its use before they began providing explanations. For rationale collection, participants engaged in a turn-taking experience where they observed an action and then explained it while the game is paused (Figure~\ref{fig:Play_game}). 
While thinking out loud, an automatic speech recognition library~\cite{github_2017} transcribed the utterances, substantially reducing participant burden and making the flow more natural than having to type down their utterances. 
Upon game play completion, the players reviewed all action-explanation pairs in a global context by replaying each action~(Figure~\ref{fig:Review_rationale}). 
We deployed our data collection pipeline on Turk Prime (a wrapper over Amazon Mechanical Turk) and collected over 2000 unconstrained action-rationale pairs from 60 participants. 
\begingroup
\setlength{\tabcolsep}{6pt} 
\renewcommand{\arraystretch}{1.2} 
\begin{table}[t]
  \vspace{-5mm}
  \caption{Examples of different rationales generated for the same game action.}
  \label{tab:examples}
  \begin{tabular}{p{0.1\columnwidth}|p{0.3\columnwidth}p{0.5\columnwidth}}
    \toprule
    {\bf Action} & {\bf Focused-view} & {\bf Complete-view}\\
    \midrule
    Right & 
    I had cars to the left and in front of me so I needed to move to the right to avoid them. & 
    I moved right to be more centered. This way I have more time to react if a car comes from either side. \\
    \midrule
    Up & 
    The path in front of me was clear so it was safe for me to move forward.  & 
    I moved forward making sure that the truck won\textquotesingle t hit me so I can move forward one spot. \\
    \midrule
    Left & 
    I move to the left so I can jump onto the next log. & 
    I moved to the left because it looks like the logs and top or not going to reach me in time, and I\textquotesingle m going to jump off if the law goes to the right of the screen. \\
  \bottomrule
  \end{tabular}
  \vspace{-5mm}
\end{table}
\endgroup
\subsubsection{Neural Model Configuration}

We use the same 
encoder-decoder RNN as in phase 1, but this time, we varied the input configurations with the intention of producing varying styles of rationales to experiment with different strategies for rationale generation. 
Last time, we deployed just one configuration, the \textit{focused-view }configuration. 
This focused-view configuration accurately reflects what the agent is considering, leading to concise rationales due to the limitation of data the agent had available for rationale generation. 
To contrast this, we formulated a second \textit{complete-view} configuration that gives the rationale generator the ability to use all information on the screen. 
We speculated that this configuration would produce more detailed, holistic rationales and use state information that the algorithm is not considering. 
See Table \ref{tab:examples} for example rationales generated by our system. 
However, it remains to be seen if these configurations produce perceptibly different rationales to users who do not have any idea of the inner workings of the neural network. 
We evaluated the alignment between the upstream algorithmic decisions and downstream user effects using the user studies described below. 
\subsubsection{Evaluation}

\begin{figure}[t]
    \centering
    \begin{minipage}[t]{0.5\textwidth}
        \centering
        \includegraphics[width=\textwidth]{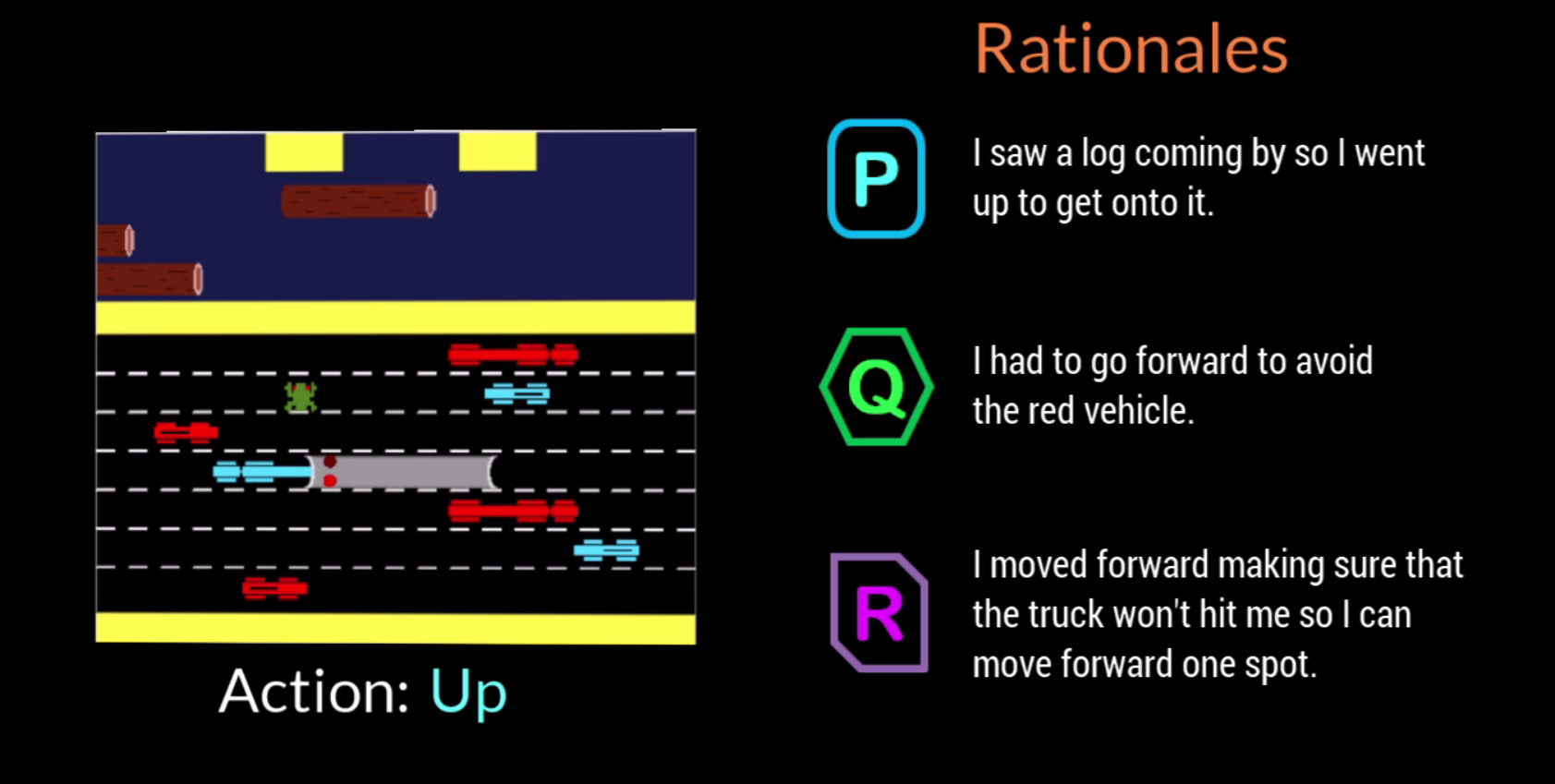} 
        \captionsetup{width=0.95\linewidth, font= small}
        \caption{User study screenshot depicting the action and the rationales: \textit{P = Random (lower baseline), Q = Exemplary (higher baseline), R = Our model (Candidate)}}
        \label{fig:Study}
    \end{minipage}\hfill
    \begin{minipage}[t]{0.45\textwidth}
        \centering
        \includegraphics[width=\textwidth]{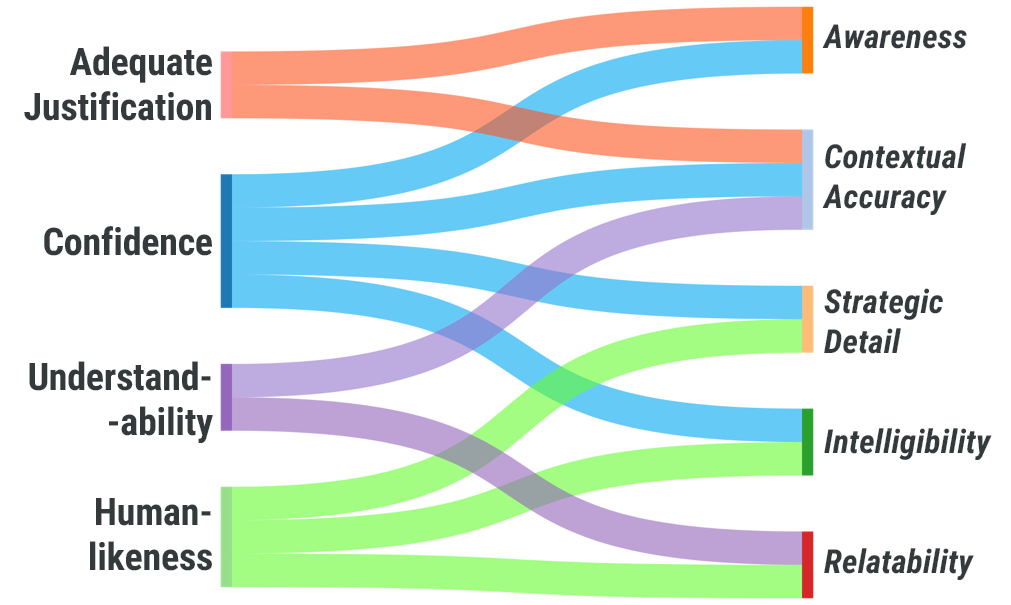} 
        \captionsetup{width=0.95\linewidth, font = small}
        \caption{Emergent relationship between the dimensions (left) and components (right) of user perceptions and preference}
        \label{fig:dim_and_comp}
    \end{minipage}
\end{figure}

For phase 2, the evaluation of the XAI system was entirely qualitative, human-based analysis. 
We conducted two user studies: the first establishes that, when compared against baselines, both network configurations produce plausible outputs; the second establishes if the outputs are indeed perceptibly different to “naïve” users who are unaware of the neural architecture and explores contextual user preferences. 
In both user studies, participants watched videos where the agent is taking a series of actions and “thinking out loud” in different styles~(see Figure \ref{fig:Study} for implementation details).

The first user study established the viability of generated rationales, situating user perception along the dimensions of \textit{confidence, human-likeness, adequate justification,} and \textit{understandability}. 
We adapted these constructs from our findings in phase 1, technology acceptance models (e.g., UTAUT)~\cite{venkatesh2003user,davis1989perceived}, and related research in HCI \cite{chernova2009confidence,kaniarasu2013robot,beer2011understanding}. 
Analyzing the qualitative data, we found emergent components that speak to each dimension; see \cite{ehsan2019automated} for details of the analysis. 
For \textit{confidence}, participants found that contextual accuracy, awareness, and strategic detail are important in order to have faith in the agent’s ability to do its task. 
Whether the generated rationales appear to be made by a human (\textit{human-likeness}) depended on their intelligibility, relatability, and strategic detail. 
In terms of \textit{explanatory power} (adequate justification), participants prefer rationales with high levels of contextual accuracy and awareness. 
For the rationales to convey the agent’s motivations and foster \textit{understandability}, they need high levels of contextual accuracy and relatability (see Figure \ref{fig:dim_and_comp} for a mapping and Table \ref{tab:components} for definitions of these components).

\begin{table}[t]
  \caption{Descriptions for the emergent \textit{components} underlying the human-factor \textit{dimensions} of the generated rationales (See \cite{ehsan2019automated} for further details).}
  \label{tab:components}
  \begin{tabular}{p{0.25\linewidth}|p{0.6\linewidth}}
    \toprule
    {\bf Component} & {\bf Description}\\
    \midrule
    \textit{Contextual Accuracy} & Accurately describes pertinent events in the context of the environment.\\
    \midrule
    \textit{Intelligibility} & Typically error-free and is coherent in terms of both grammar and sentence structure.\\
    \midrule
    \textit{Awareness} & Depicts and adequate understanding of the rules of the environment.\\
    \midrule
    \textit{Relatability} & Expresses the justification of the action in a relatable manner and style.\\
    \midrule
    \textit{Strategic Detail} & Exhibits strategic thinking, foresight, and planning.\\
  \bottomrule
  \end{tabular}
\end{table}
In the second user study, we found that there is alignment between the intended differences in features of the generated rationales and the perceived differences by users. 
Without any knowledge beyond what is shown on the video, they described the difference in the styles of the rationales in a way that was consistent with the intended differences between them. This finding is an important secondary validation of how upstream algorithmic changes in neural network configuration lead to the desired user effects downstream. 

The second user study also explores user preferences between the focused-view and complete-view rationales along three dimensions:  \textit{confidence} in the autonomous agent, communication of \textit{failure} and \textit{unexpected behavior}. 
We found that, context permitting, participants preferred detailed rationales so that they can form a stable mental model of the agent’s behavior.
\begingroup
\setlength{\tabcolsep}{8pt} 
\renewcommand{\arraystretch}{1} 
\begin{table}[t]
  \vspace{-5mm}
  \caption{Side-by-side comparison of each phase in the case study}
  \label{tab:two-phase-comparison}
  \begin{tabular}{p{0.2\columnwidth}|p{0.35\columnwidth}|p{0.35\columnwidth}}
    \toprule
    {\bf } & {\bf Phase 1} & {\bf Phase 2}\\
    \midrule
    \textit{Data Collection} & 
    225 action-rationale annotations from 12 people &
    Over 2000 action-rationale annotations from 60 people \\
    \midrule
    \textit{Corpus} & 
    Semi-synthetic grammar on top of natural language &
    Fully unconstrained natural language; no grammar \\
    \midrule
    \textit{Neural Network Configuration} & 
    Only one setup: focused-view, a $7\times7$ window around the agent &
    Two configurations: focused-view and complete-view designed to produce concise vs. detailed rationales \\
    \midrule
    \textit{Evaluation} &
      Part procedural; part human-based evaluation along one dimension – satisfaction of explanation &
      Full human-based evaluation with metrics defining plausibility against baselines using two studies \\
      \midrule
    \textit{Key Lessons} &
      The technique works to produce accurate rationales that are satisfactory to humans. User study insights help unpack what it means to be “satisfactory”, which enables the next generation of systems in Phase 2. &
      Both configurations produce plausible rationales that are perceptibly different to end-users. User studies further reveal underlying components of user perceptions and preferences, refining our understanding of \textit{“who”} the human is.\\
  \bottomrule
  \end{tabular}
  \vspace{-5mm}
\end{table}
\endgroup
\subsection{Summary}

As we wrap up our case study overview, we want to underscore how technology development and understanding of human factors co-evolve together. 
In the following section, we will see how the foundation laid by the case study generates new areas of research, enabling a \textit{``turn to the sociotechnical''} for the HCXAI paradigm. 
\section{What's Next: Turn to the Sociotechnical}
At first glance, it may appear that a case-study using {\em Frogger} is not representative of a real-world XAI system.
However, therein lies a deeper point---considering issues of fairness, accountability, and transparency of sociotechnical systems, it is risky to directly test out these systems in mission-critical domains without a formative and substantive understanding of the human factors around XAI systems. 
By conducting the case study in a controlled setting as a first step, we obtain a formative understanding of the technical and human sides, which can then be utilized to better implement such systems in the wild. 
Subsequent empirical and theoretical work can then build on any transferable insights from this work. 

Building on our insights, we will outline two areas of investigation and share preliminary challenges and opportunities:
$(1)$~Perception differences due to users' backgrounds, and 
$(2)$~Social signals and explanations.
These areas are by no means exhaustive; rather, these are ones that have come to light from our case study. It's important to note here that, without the formative insights from multiple phases of our case study, the depth and richness of the research areas would not have been obvious. 
That is, while we considered multiple end-users (developers, non AI-experts, etc.), the case study's findings highlighted further non-obvious striations in the technical and social aspects of human perceptions of XAI.

\subsection{Perception differences due to users' backgrounds}

\textit{How do people of different professional and epistemic backgrounds perceive the same XAI system? Do their backgrounds impact their perception?} 
These questions came from the observation that explanations, by definition, are context-sensitive. 
The \textit{who} governs \textit{how} the \textit{why} is most effectively conveyed. 
Moreover, qualitative data analysis in our case study also hinted that people’s professional and educational backgrounds impact their perception of explanations. %
The differences in perception were salient namely in the dimensions of \textit{confidence} and \textit{understandability}.
These differences were particularly remarkable for people who were familiar with the technical side of computing compared to ones who were not.
This observation sparked the question: what might be the different explainability needs for end-users with different backgrounds? How might we go about teasing this apart? 

From a methodological standpoint, we can run user studies similar to that in our case studies to get a formative understanding of how backgrounds impact perception and preferences of XAI systems. 
For instance, we can provide the same explanation to two related yet different groups (e.g., engineers vs. lay-riders of self-driving cars) and investigate if and how their backgrounds impact perception and preferences. 
\subsection{Social signals and explanations}
\textit{What roles might social signals, especially in a team-based collaborative setting, play in HCXAI? How might we embed social transparency into our systems in order to facilitate user actions?}
This research interest stems from the observation that we seldom find consequential AI systems in isolated settings where only one human interacts with the machine. 
Rather, most systems are socially situated in organizational settings involving teams of people engaging in collaborative decision-making. 
How will our design evolve as we move beyond the 1-1 human-computer interaction paradigm?
When we talk about a paradigm beyond the 1-1 human-computer interaction, we are referring to situations where the collaborative decision-making and relationships of multiple individuals in an organization or a team are mediated through technology. The scenario is complex now because we have two types of relationships to consider: the type that is between the machine and the humans as well as the interdependent accountability amongst different kinds of stakeholders.

Let us consider the following scenario:
In an IT setting, Cloud Solutions architects often need to make purchasing decisions around Virtual Machine (VM) instances that help the organization run online, mission-critical services on the cloud. 
There are real costs of ``wrong-sizing'' the VM instance---if you underestimate, the company's system might become overloaded and crash; if you overestimate, the company wastes valuable monetary resources.
Moreover, there are teams of people who are secondary and tertiary stakeholders of the VM instances.
Suppose an AI system recommends certain parameters for the VM instances to a single Solutions architect who is accountable to and responsible for the other stakeholders. 
The AI system also provides ``technical'' explanations by contextualizing the recommendation with past usage data analytics.
Given the interpersonal and professional accountability risks, is technical explainability enough to give the engineer the confidence to accept the AI's recommendation? 
Or does the explanation need to incorporate the embedded, interconnected nature of stakeholders such as the use of social signals? 
Social signals here can be thought of as digital footprints that provide context of the team's perspective on the collaborative decision-making; for instance, stakeholders can give a “+1” or an upvote on the recommendation. 

From a methodological perspective, we can design between-subject user studies where we measure the perceptions of collaborative decision-making. 
One group would only get technical explanations while the other group gets both social and technical signals. 
We can simulate the aforementioned scenario and measure how confident each group is in their decisions to act on the right-sizing recommendations.  

\subsection{Summary: Socially Situated XAI Systems}
In considering these research directions, we should appreciate the value of controlled user studies in generating formative insights. 
However, if we ignore the socially situated nature of our technical systems, we will only get a partial, unsatisfying picture of the \textit{“who”}.  
Therefore, enhancing the current paradigm with sociotechnical approaches is a necessary step. 
This is because consequential technological systems are almost always embedded in a rich tapestry of social relationships. 
Take, for example, the aforementioned scenario with right-sizing VM instances. 
Our on-going work has shown that the organizational culture and its perception of AI-systems strongly impacts people’s confidence to act on machine-driven recommendations, no matter how technically explainable they are. 
Any organizational environment carries its own socio-political assumptions and biases that influence technology use~\cite{suchman2007human}. 
Understanding the rich social factors surrounding the technical system may be as equally important to the adoption of explanation technologies as the technology itself.
Designing for the sociotechnical dynamics will require us to understand the rich, contextual, human experience where meaning-making is constructed at the point of interaction between the human and the machine. 
But how might we go about it? 
We will need to think of ways to critically reflect on methodological and conceptual challenges. 
In the following section, we lay out some strategies to handle these conceptual blocks.

\section{Human-centered XAI, Critical Technical Practice, and the Sociotechnical Lens}

The prior section highlights the socially situated nature of XAI systems that demand a sociotechnical approach of analysis. 
With each hypothesis and technical advancement, the resolution of \textit{"who" the human is} improved.
As the metaphorical picture of the user became clearer, other people and objects in the background have also come into perspective. 
The newfound perspective demands the ability to incorporate all parties into the picture.
It also informs the technological development needs of the next generation of refinement in our understanding of the \textit{"who"}.
As the domain of HCXAI evolves, so must our epistemological stances and methodological approaches. 
Currently, there is not a singular path to construct the sociotechnical lens and nor should there be given the complexity and richness of human connections. 
However, we have a rich foundation of prior work both in AI and HCI that will help us get there. 
In developing the sociotechnical lens of HCXAI, we are particularly inspired by prior work from Sengers et al.~\cite{sengers2005reflective}, Dourish et al.~\cite{dourish2004action,dourish2004reflective}, and Friedman et al.~\cite{friedman2008value} 

In particular, we believe that viewing HCXAI through the perspective of a Critical Technical Practice (CTP) will foster the grounds for a {\em reflective} HCXAI. 
CTP~\cite{agre1997toward,agre1997computation} encourages us to question the core assumptions and metaphors of a field of practice, critically \textit{reflect} on them to overcome impasses, and generate new questions and hypotheses. 
By {\em reflection}, we refer to ``\textit{critical} reflection [that] brings unconscious aspects of experience to conscious awareness, thereby making them available for conscious choice''~\cite{sengers2005reflective}.

Our perspective on reflection is grounded in critical theory~\cite{held1980introduction,feenberg1991critical} and inspired by Sengers et al.'s notion of Reflective Design~\cite{sengers2005reflective}. 
We recognize that the lens through which we look at and reason about the world is shaped by our conscious and, more importantly, unconscious values and assumptions.
These values, in turn, become embedded into the lens of our technological practices and design.
By bringing the unconscious experience to our conscious awareness, critical reflection not only allows us to look \textit{through} the lens, but also \textit{at} it.
A reflective HCXAI creates the necessary intellectual space to make progress through conceptual and technical impasses while the metamorphosis of the field takes place.
Given that the story of XAI has just begun, it would be premature to attempt a full treatise of human-centered XAI. 
However, we can begin with two key properties of a reflective HCXAI: 
$(1)$~a domain that is \textit{critically reflective} of (implicit) assumptions and practices of the field, and 
$(2)$~one that is \textit{value-sensitive} to both users and designers.  

In the rest of this section, we will provide relevant background about CTP, how it allows HCXAI to be reflective, why it is useful, and complimentary strategies from related fields that can help us build the sociotechnical lens. 
We will also contextualize the theoretical proposal with a scenario and share the affordances in explainability we gain by viewing HCXAI as a Critical Technical Practice. 
We conclude the section with challenges of a reflective HCXAI.

\subsection{Reflective HCXAI using a Critical Technical Practice Lens} 
The notion of Critical Technical Practice was pioneered by AI researcher Phil Agre in his 1997 book, {\em Computation and Human Experience}~\cite{agre1997computation}. 
CTP encourages us to question the core assumptions and metaphors of a field and critically reflect on them in order to overcome impasses in that field. 
In short, there are four main components of the perspective: 
(\RNum{1})~identify the core metaphors and assumptions of the field, 
(\RNum{2})~notice what aspects become marginalized when working within those assumptions, 
(\RNum{3})~bring the marginalized aspects to the center of attention, and 
(\RNum{4})~develop technology and practices to embody the previously-marginalized components as alternative technology.
Using the CTP perspective, Agre critiqued the dominant narrative in AI at the time, namely abstract models of cognition, and brought situated embodiment central to AI's perspective on intelligence. 
By challenging the core metaphor, they successfully opened a space for AI that led to advancements in the new "situated action" paradigm~\cite{suchman2007human}.

In our case, we can use the CTP perspective to reflect on and question some of the dominant metaphors in Explainable AI.
This reflection can expand our design space by helping us identify aspects that have been marginalized or overlooked.
For instance, one of the dominant narratives in XAI makes it appear as though interpretability and explainability are model-centered problems, which is where a lot of current attention is rightfully invested. 
However, our experiences while broadening the lens of XAI has led us to reflect on explainability, leading to an important question: 
where does the \textit{``ability''} in explain-ability lie?
Is it a property of the model or of the human interpreting it, or is it a combination of the two? 
What if we switch the `\textit{`ability''} in interpretability or explainability to the human? 
Or perhaps there is a middle ground where meaning is co-creatively manifested at the point of action between the machine and the human? 
By enabling critical reflections on core assumptions and impulses in the field, the CTP perspective can be the lighthouse
that guides us as we embark on a reflective HCXAI journey and navigate through the design space.

There are three main affordances of the CTP approach in HCXAI. 
First, the perspective allows traversal of the marginalized insights ---in this case the human-centered side of XAI---to come to the center, which can open new design areas previously undetected or under-explored. 
Second, the critical reflection mindset can enable designers to think of new ways to understand human factors. 
It can also empower users with new interaction capabilities that promote their voices in technologies, which, in turn, can improve our understanding of \textit{``who'' the human is} in HCXAI.
Take, for instance, our understanding of user trust. 
To foster trust, a common impulse is to aim for the "positive" direction and nudge the human to find the machine's explanation plausible and to accept it.
As our case study shows, this is certainly a viable route.
However, should that be the only route? 
That is, should this impulse for user-agreeableness be the only way to understand this human factor of trust?
In certain contexts, like fake news detection, might we be better off by designing to evoke reasonable skepticism and critical reflection in the user? 
Since no model is perfect at all times, we cannot expect generated explanations to always be correct.
Thus, creating the space for users to voice their skepticism or disagreement not only empowers new forms of interaction, but also allows the user to become sensitive to the limitations of AI systems. 
Expanding the ways we reason about fostering trust can create a design perspective that is not only reflective but is also pragmatic.
Third, critical reflection can help us defamiliarize and decolonize our thinking from the dominant narratives, helping us to not only look \textit{``through''} but also \textit{``at''} the sociotechnical lens of analysis. 

\subsection{Strategies to Operationalize Critical Technical Practice in HCXAI }
To operationalize the CTP perspective, we can incorporate rich strategies from other methodological traditions rooted in HCI, critical studies, and philosophy such as participatory design~\cite{bodker1991through,ehn1992scandinavian}, value-sensitive design~\cite{friedman2008value,friedman2012envisioning}, reflection-in-action~\cite{schon2017reflective,wright2004technology,djajadiningrat2000interaction}, and ludic design~\cite{gaver2000alternatives,gaver2004drift}. 
Reflective HCXAI does not take a normative stance to privilege one design tradition over the other, nor does it replace one with the other; rather, it incorporates and integrates insights and methods from related disciplines. 
For our current scope, we will briefly elaborate on two approaches---{\em participatory design} (PD) and {\em value-sensitive design} (VSD). 


Participatory Design  challenges the power dynamics between the designer and user and aims to support democratic values at every stage of the design process. 
Not only does it advocate for changing the system but also challenges the practices of design and building, which might help bring the marginalized perspectives to the forefront. 
This fits in nicely with one of the key properties of a reflective HCXAI: the ability to critically reflect on core assumptions and politics of both the designer and the user. 


Value-Sensitive Design  is ``a theoretically grounded approach to the design of technology that seeks to account for human values in a principled and comprehensive manner throughout the design process”~\cite{friedman2008value}. 
Using {\em Envisioning cards}~\cite{friedman2012envisioning}, researchers can engage in exercises with stakeholders to understand stakeholder values, tensions, and political realities of system design. 
A sociotechnical approach by construction, it incorporates a mixture of conceptual, empirical, and technical investigations stemming from moral philosophy, social-sciences, and HCI. We can use it to investigate the links between the technological practices and values of the stakeholders involved. 
VSD aligns well with the other key property of reflective HCXAI: being value-sensitive to both designers and users.


With the theoretical and conceptual blocks in mind, let us look at a scenario that might help us contextualize the role of the CTP perspective, VSD, and PD in a reflective HCXAI paradigm. 
This scenario is partially-inspired by our on-going work with teams of radiologists. 
In a large medical hospital in the US, teams of radiologists use an AI-mediated task list that automatically prioritizes the order in which radiologists go through cases (or studies) during their shifts. 
While a prioritization task might seem trivial at first glance, this one has real consequences---failure to appropriately prioritize has consequences ranging from a missed report deadline to ignoring an emergency trauma patient. 

The CTP perspective encourages us to look at the dominant narrative and think of marginalized perspectives to expand our design space. 
Here, we should critically reflect on the role of explanations in this system. In such a consequential system, fostering user trust is a core goal. 
Considering that the AI model might fail, is trust best established by creating explanations that always nudge users to accept the AI system's task prioritization? 
Or might we design with the goal of user reflection instead of user acceptance? 
Reflection can be in the form of reasonable skepticism. 
In fact, skepticism and trust go hand in hand; skepticism is part of that critical reflective process that helps us question our core assumptions. 
Even if we could build such a system, how might we \textit{evaluate} explanations that foster reflection instead of acceptance? What type of prioritization tasks should privilege acceptance vs. reflection? 

The answers to these questions are not apparent without a sociotechnical approach and constructive engagement with the communities in question. 
Having identified some of the marginalized aspects and critically reflecting on them using the CTP perspective, we can use the aforementioned strategies, such as participatory design (PD) and value-sensitive design (VSD), to operationalize the reflective HCXAI perspective. 
For instance, we can use the PD approach to ensure the power dynamics between designers and users are democratic in nature. 
Moreover, we can reflexively recognize the politics of the design practice and reflect on how we build any interventions. 
We can also incorporate VSD elicitation exercises using the Envisioning Cards to uncover value tensions and political realities in the hospital systems. 
For instance, what, if any, are tensions between the values of the administration, the insurance industry, and the radiologists? 
What values do the different stakeholders feel the XAI system should embody and how do these values play off of each other in terms of alignment or tensions? 

\subsection{Challenges of a reflective HCXAI paradigm}
With the affordances of a reflective HCXAI in mind, we observe two current challenges where we need a concerted community effort. %
First, sociotechnical work requires constructive engagement with partner communities of practice. 
Our end-users live in communities of practices that have their own norms (e.g., radiologists within the community of medical practice). 
As outsiders, we cannot expect to gain an embedded understanding of the “who” without constructively engaging with partner communities (e.g, radiologists) on their own terms and timelines. 
This means we need to be sensitive to their values as well as norms to foster sustainable community relationships. 
Not only are these endeavors resource and time intensive, which could impact publication cycles, but they also require stakeholder buy-in at multiple levels across organizations. 

Second, sociotechnical work in a reflective HCXAI paradigm would require active translational work from a diverse set of practitioners and researchers. 
This entails that, compared to \boldmath$\top$-shaped researchers who have intellectual depth in one area, we need more  $\rm{\Pi}$-shaped ones who have depth in two (or more) areas and thus the ability to bridge the domains. 
\section{Conclusions}

As the field of XAI evolves, we recognize the socially situated nature of consequential AI systems and re-center our focus on the human.
We introduce Human-centered Explainable AI (HCXAI) as an approach that puts the human at the center of technology design and develops a holistic understanding of \textit{``who''} the human is.
It considers the interplay of values, interpersonal dynamics, and socially situated nature of AI systems. 
In particular, we advocate for a reflective sociotechnical approach that incorporates both social and technical elements in our design space.
Using our case study that pioneered the notion of \textit{rationale generation}, we show how technical advancements and the understanding of human factors co-evolve together.
We outline open research questions that build on our case study and highlight the need for a reflective sociotechnical approach.
Going farther,
we propose that a  reflective HCXAI paradigm---using the perspective of Critical Technical Practice and strategies such as participatory design and value-sensitive design---will not only help us question the dominant metaphors in XAI, but they can also open up new research and design spaces.
\section*{Acknowledgements}
Sincerest thanks to all past and present teammates of the Human-centered XAI group at the Entertainment Intelligence Lab whose hard work made the case study possible---Brent Harrison, Pradyumna Tambwekar, Larry Chan, Chenhann Gan, and Jiahong Sun.
Special thanks to Dr. Judy Gichoya for her informed perspectives on the medical scenarios. 
We'd also like to thank Ishtiaque Ahmed, Malte Jung, Samir Passi, and Phoebe Sengers for conversations throughout the years that have constructively added to the notion of a `Reflective HCXAI'. 
We are indebted to Rachel Urban and Lara J. Martin for their amazing proofreading assistance.
We are grateful to reviewers for their useful comments and critique.
This material is based upon work supported by the National Science Foundation under Grant No. 1928586.

\balance{}

\bibliographystyle{splncs04}
\bibliography{sample-bib}
%

\end{document}